\documentclass{elsart}

\usepackage{graphicx}
\usepackage{amssymb}

\def\NIMA#1#2#3{Nucl. Instr. Meth. Phys. Res. A {\bf #1}\ (#2)\ #3}

\def\IEEE#1#2#3{IEEE Trans. Nucl. Sci. vol. {\bf #1}\ (#2)\ #3}

\begin{document}

\begin{frontmatter}

\title{Pulse height measurements and electron attachment in drift chambers 
operated with Xe,CO$_2$ mixtures}

\author[gsi]{A.~Andronic\thanksref{info}},
\author[hei]{H.~Appelsh\"auser}, 
\author[gsi]{C.~Blume}, 
\author[gsi]{P.~Braun-Munzinger}, 
\author[mue]{D.~Bucher}, 
\author[gsi]{O.~Busch}, 
\author[gsi]{A.~Castillo Ramirez}, 
\author[buc]{V.~C\u at\u anescu}, 
\author[buc]{M.~Ciobanu}, 
\author[gsi]{H.~Daues}, 
\author[gsi]{A.~Devismes}, 
\author[hei]{D.~Emschermann}, 
\author[dub]{O.~Fateev},
\author[gsi]{C.~Garabatos}, 
\author[hei]{N.~Herrmann}, 
\author[gsi]{M.~Ivanov}, 
\author[hei]{T.~Mahmoud}, 
\author[mue]{T.~Peitzmann}, 
\author[hei]{V.~Petracek}, 
\author[buc]{M.~Petrovici},
\author[mue]{K.~Reygers}, 
\author[gsi]{H.~Sann}, 
\author[mue]{R.~Santo},
\author[hei]{R.~Schicker},  
\author[gsi]{S.~Sedykh}, 
\author[dub]{S.~Shimansky},
\author[gsi]{R.S.~Simon}, 
\author[dub]{L.~Smykov},
\author[hei]{H.K.~Soltveit}, 
\author[hei]{J.~Stachel}, 
\author[gsi]{H.~Stelzer}, 
\author[gsi]{G.~Tsiledakis}, 
\author[hei]{B.~Vulpescu}, 
\author[gsi]{J.P.~Wessels}, 
\author[hei]{B.~Windelband},
\author[mue]{O.~Winkelmann},
\author[hei]{C.~Xu},
\author[mue]{O.~Zaudtke},
\author[dub]{Yu.~Zanevsky},
\author[mue]{V.~Yurevich}

\address[gsi]{Gesellschaft f\"ur Schwerionenforschung, Darmstadt, Germany}
\address[hei]{Physikaliches Institut der Universit\"at Heidelberg, Germany}  
\address[mue]{Institut f\"ur Kernphysik, Universit\"at M\"unster, Germany}  
\address[buc]{NIPNE Bucharest, Romania}  
\address[dub]{JINR Dubna, Russia}  

{for the ALICE Collaboration}

\thanks[info]{Corresponding author: GSI, Planckstr. 1, 64291 Darmstadt,
Germany; Email:~A.Andronic@gsi.de; Phone: +49 615971 2769; 
Fax: +49 615971 2989}

\begin{abstract}
We present pulse height measurements in drift chambers operated with Xe,CO$_2$ 
gas mixtures. 
We investigate the attachment of primary electrons on oxygen and SF$_6$ 
contaminants in the detection gas.
The measurements are compared with simulations of properties of drifting 
electrons.
We present two methods to check the gas quality: gas chromatography and
$^{55}$Fe pulse height measurements using monitor detectors.
\end{abstract}

\end{frontmatter}

\section{Introduction} \label{aa:intro}

The ALICE Transition Radiation Detector (TRD) \cite{aa:tdr} has to provide 
both electron identification and particle tracking. To achieve this, 
accurate pulse height measurement in drift chambers operated with 
Xe,CO$_2$(15\%) gas mixture over the drift time of the order of 2~$\mu$s 
(spanning 3~cm of drift length) is a necessary requirement.
For such precision measurements, it is of particular
importance not to loose charge by electron attachment, i.e. the absorption of 
drifting electrons by electronegative molecules present in the detector gas 
as contaminants.
The large volume (28~m$^3$) of the ALICE TRD and the high cost of xenon make the 
above arguments very serious for the operation of the final detector.

Attachment is a well studied and generally understood phenomenon, both
fundamentally \cite{aa:chris2,aa:chris} and concerning its practical implications
for gas drift chambers \cite{aa:huk}.
For electron energies relevant to gaseous detectors (energies below a few eV), 
attachment occurs mainly via two mechanisms: resonance capture and dissociative 
capture. Resonance capture, also called Bloch-Bradbury mechanism \cite{aa:bb}, 
has the largest cross-section. It can be written as: 
\begin{equation} 
I + e^- \rightarrow I^{-*} \label{aa:eq1}
\end{equation}
\begin{equation} 
I^{-*} + S \rightarrow I^- + S^* . \label{aa:eq2}
\end{equation}
$I$ denotes the impurity and $S$ is a third body stabilizer, which 
in case of gas detectors, is usually the quencher. 
The star ($*$) denotes a vibrationally excited state. Besides decaying by the 
resonant energy transfer (\ref{aa:eq2}), $I^{-*}$ could also decay by electron 
emission (autodetachment), in which case there is no signal loss. 
The rate of process (\ref{aa:eq2}), and thus the magnitude of attachment,
depends on the concentration of $S$ and on the lifetime of the 
excited state $I^{-*}$.
Also, as a result of different vibrational levels available for the energy
transfer (\ref{aa:eq2}), the attachment depends on the type of quencher 
\cite{aa:huk,aa:wen}.
An excess of electron attachment with respect to the Bloch-Bradbury mechanism
has been identified and assigned to van der Waals complexes \cite{aa:kok}.

The most common electronegative molecule is O$_2$,
present in gaseous detectors as a residual contaminant in the gas supply or from 
the atmosphere due to imperfect tightness of the system.
Attachment on O$_2$ has been extensively studied for Ar-based mixtures in ranges 
of parameters relevant for drift chambers \cite{aa:huk,aa:wen}.
Another common contaminant in gas detectors is H$_2$O, usually outgassed by 
assembly materials, and often appreciated as an ageing-limiting agent 
\cite{aa:kad}.
It has been found that, for certain Ar-based mixtures, attachment on H$_2$O 
alone is negligible, but a few hundred ppm of H$_2$O can double the attachment 
coefficient on O$_2$ \cite{aa:huk}.

In one of our early measurements with TRD prototype chambers operated 
with Xe,CH$_4$(10\%) we observed electron attachment under very low contamination
levels of O$_2$ and H$_2$O. 
As that particular gas supply was exhausted during the measurements, we were not 
able to analyze it and so not able to attribute the attachment to a defined 
impurity.
Subsequent measurements (performed with a new supply of Xe) proved to be free 
of visible attachment.
However, recent 
observation of strong attachment, this time with the mixture Xe,CO$_2$(15\%) 
forced us to investigate the issue in more detail.
We have been able to identify the impurity responsible for attachment:
sulphur hexafluoride, SF$_6$, which was found to be present at the ppm level in 
a xenon supply.
This heavy gas, well known for its excellent insulating properties,
has an extremely large cross section for electron attachment at low electron 
energies.
We note that pulse height distributions as a function of drift time were 
measured before in drift chambers with Xe-based mixtures, also related to TRDs
\cite{aa:wat,aa:zeus,aa:hol,aa:d0}. 
A decrease of the average pulse height as a function of drift time was observed
in all these cases and it was attributed to electron attachment \cite{aa:d0},
but not quantitatively understood.

Here we report on measurements performed during prototype tests of the
ALICE TRD \cite{aa:tdr}.
Drift chambers operated with Xe,CO$_2$ mixtures are investigated.
The experimental setup and method of data analysis are described in the next 
section. We then present measurements under clean conditions (no attachment).
The following sections contain our measurements of attachment on oxygen and
on SF$_6$. We compare our results with simulations of charge transport
for various gas mixtures.  
We present gas quality checks employing gas chromatograph analyses and 
$^{55}$Fe measurements using especially-built monitoring detectors.

\section{Experimental setup} \label{aa:meth}

\begin{figure}[hbt]
\centering\includegraphics[width=.6\textwidth]{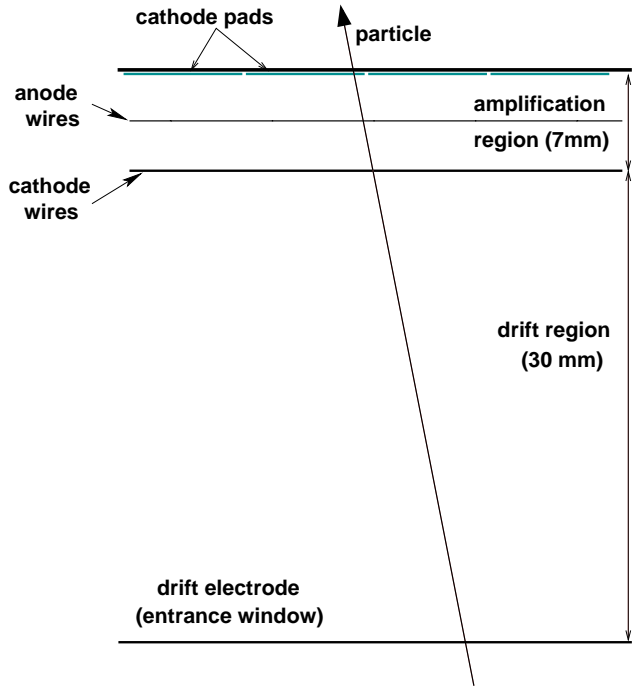}
\caption{Schematics of the drift chamber.}
\label{aa:prin} 
\end{figure}

Most of the results are obtained using prototype drift chambers (DC) with 
a construction similar to that anticipated for the final ALICE TRD
\cite{aa:tdr}, but with a smaller active area (25$\times$32~cm$^2$).
In Fig.~\ref{aa:prin} we present a schematic view of the detector.
The DC has a drift region of 30~mm and an amplification region of 7~mm.
The anode wires (W-Au, 20~$\mu$m diameter) have a pitch of 5~mm.
For the cathode wires (Cu-Be, 75~$\mu$m diameter) we use a pitch of 2.5~mm.
We read out the signal on a segmented cathode plane.
The pads (of 6~cm$^2$ each) have either chevron \cite{aa:yu} or rectangular
shape.
The entrance window (25~$\mu$m aluminized Kapton) simultaneously serves  
as gas barrier and as drift electrode.

A charge-sensitive preamplifier/shaper (PASA) was especially designed and built 
(with discrete components) for the purpose of prototype tests.
It has a gain of 2~mV/fC and noise of about 1800 electrons r.m.s. 
The FWHM of the output pulse is about 100~ns for an input step function. 
For the read-out of the DC we use an 8-bit non-linear Flash ADC (FADC) system 
with 100~MHz sampling frequency, 0.6~V voltage swing and adjustable baseline.
The FADC sampling was rebinned in the off-line analysis in order to resemble
the performance of the final detector \cite{aa:tdr}.
The data acquisition (DAQ) is based on a VME event builder and was developed 
at GSI Darmstadt \cite{aa:mbs}.
As the beam diameter is of the order of a few cm, we usually limit the readout 
of the DC to 8 pads. This also minimizes data transfer on the VSB bus connecting 
the FADC and the event builder.

The measurements were carried out at beam momenta of 1~GeV/c at GSI 
Darmstadt \cite{aa:gsipi} and 3~GeV/c at the CERN PS \cite{aa:cernpi}.
The beams were mixtures of electrons and negative pions.
For the present analysis we have selected clean samples of pions using 
coincident upper thresholds on a Cherenkov detector and on a lead glass 
calorimeter (see ref. \cite{aa:andr} for details).
To minimize the effect of space charge on the pulse height measurement, 
which occurs for tracks at normal incidence to the anode wires (for which all 
charge collection takes place at a narrow spot on the anode wire), we adjusted 
the angle of incidence of the beam to about 15$^\circ$ with respect to the 
normal incidence to the anode wires. 
A particle trajectory through the detector is sketched in Fig.~\ref{aa:prin}.

The standard gas mixture for our detectors is Xe,CO$_2$(15\%) at 1 mbar above
atmospheric pressure.
The continuous flow of gas through the detectors is either vented out 
or recirculated via our gas system.
The detectors are usually operated at gas gains around 8000.
Our standard supply of xenon is from Messer-Griesheim \cite{aa:messer} and 
proved to provide a very good detector performance. 
We have also used xenon from Linde \cite{aa:linde}, which provided a
strikingly poor signal at first examination.
SF$_6$ was detected early on with gas chromatography techniques in 
this xenon supply, and its concentration was measured to be of 
the order of 1~ppm.
Both xenon supplies were used in beam measurements reported on below.

\section{Measurements under clean conditions} \label{aa:clean}

For these measurements we use SF$_6$-free xenon. 
The oxygen content in the gas was continuously monitored and kept below 
10~ppm using a flow of 2-3~liters of fresh gas per hour into two chambers
of about 9 liters total volume. The water content was about 150~ppm.

\begin{figure}[hbt]
\centering\includegraphics[width=.6\textwidth]{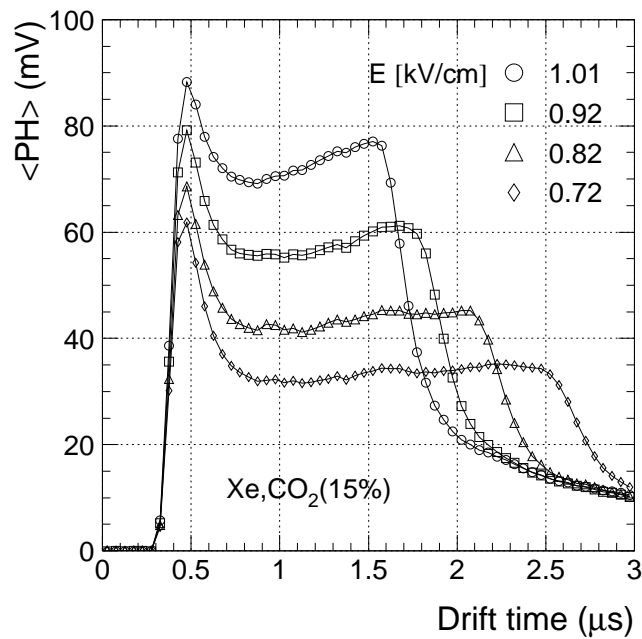}
\caption{The average pulse height as function of drift time for different
drift fields for the mixture Xe,CO$_2$(15\%).}
\label{aa:ph1} 
\end{figure}

Distributions of average pulse height, $\langle PH \rangle$, as a function 
of drift time for different drift voltages are shown in Fig.~\ref{aa:ph1}
for pions of 1~GeV/c momentum. 
The detector gas is our standard mixture, Xe,CO$_2$(15\%).
The time zero has been arbitrarily shifted by about 0.3~$\mu$s to
have a measurement of the baseline and of noise.
Similar distributions have been measured for the mixture Xe,CO$_2$(20\%).

The primary electrons from ionization energy loss of pions drift towards
the anode wire, where they are amplified. 
The signal (charge induced on the pads) is determined mainly by the 
slow-moving ions, producing long tails in the PASA output.
The overlap of these tails, convoluted with the response of the preamplifier, 
results in a slightly rising average pulse height as a function of the drift 
time, as seen in Fig.~\ref{aa:ph1}.
The peak at short drift times 
originates from the primary clusters generated in the amplification region, 
where the ionization from both sides of the anode wires contributes to the 
same time interval.
Note that, for the present conditions, lower values of drift field imply 
smaller drift velocity (see Fig.~\ref{aa:vdrift2} in section \ref{aa:calc}), 
leading to a stretching of the signal over longer drift times.
Our measurements established for the first time \cite{aa:andr} the expected 
time evolution of the signal in drift chambers of the type studied here.

\section{Attachment on oxygen} \label{aa:oxygen}

\begin{figure}[hbt]
\centering\includegraphics[width=.6\textwidth]{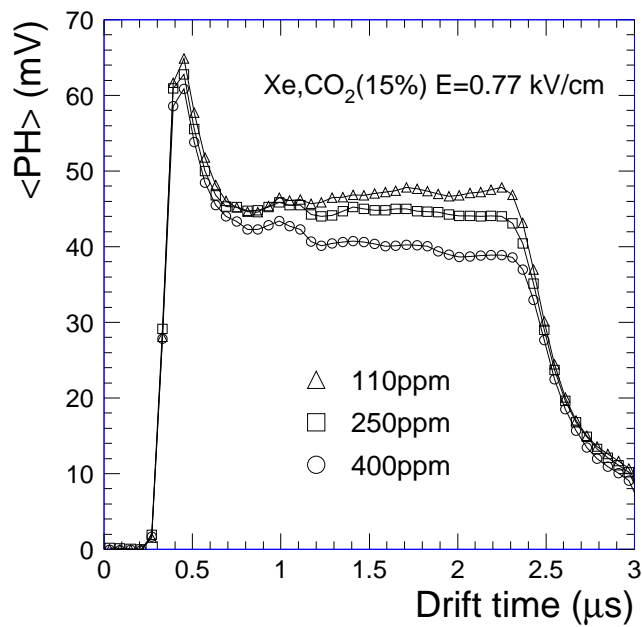}
\caption{Average pulse height as function of drift time for different
values of the oxygen concentration in Xe,CO$_2$(15\%).}
\label{aa:ph2} 
\end{figure}

Again, for these measurements we use clean (SF$_6$-free) xenon. 
In Fig.~\ref{aa:ph2} we show the average pulse height distributions as
a function of drift time for different values of the oxygen content in the
range of a few hundred ppm.
A decrease of the signal as a function of drift time is seen when the 
concentration of oxygen increases.
This is a clear indication of electron attachment. 
Notice that the signal in the amplification region, where electrons drift 
very little, is affected to a much smaller extent. 

In case of attachment, the number of electrons (and the corresponding measured
pulse height) decreases exponentially as a function of drift time, $t$:
\begin{equation} 
N(t)=N(0) \cdot \mathrm{e}^{-A\cdot t} ,
\end{equation} 
where $A$ is the attachment rate \cite{aa:huk}.
At a given gas pressure, $p$ (which is the atmospheric 
pressure in our case), the attachment rate depends linearly on the 
concentration (partial pressure) of the impurity responsible for attachment, 
$p_I$, and can be factorized \cite{aa:huk} as:
\begin{equation} 
A=p \cdot p_I \cdot C_{I},
\end{equation} 
where $C_{I}$ is the attachment coefficient of the impurity $I$.

From the above measurements of pulse height distributions as a function of drift 
time we deduce, for the present value of the drift field of 0.77 kV/cm, 
an attachment coefficient on O$_2$, $C_{O_2}$=400~bar$^{-2}\mu$s$^{-1}$.
This value is very similar to values measured for Ar,CO$_2$ mixtures with
comparable CO$_2$ content \cite{aa:wen} and more than an order or magnitude 
larger than values measured for Ar,CH$_4$ mixtures \cite{aa:huk}.
Given the short drift time in our detectors, attachment on oxygen does not
impose any severe constraint on the tightness of the drift chambers for the
final detector. 

\section{Attachment on SF$_6$} \label{aa:sf6}

All the measurements presented below have been carried out with pions of 
3~GeV/c momentum, using Linde xenon with 1.1~ppm SF$_6$ contamination.
For these measurements the O$_2$ and H$_2$O contamination was 150~ppm 
and 400~ppm, respectively.
These rather large values arise because, due to the contamination of the xenon, 
the gas was vented and not recirculated (and, as a consequence, not filtered),
as usual, through our gas system.

\begin{figure}[hbt]
\centering\includegraphics[width=.6\textwidth]{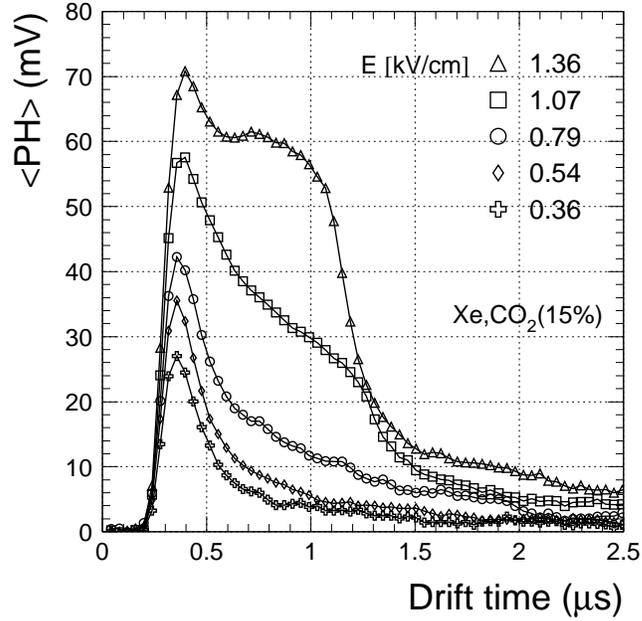}
\caption{Average pulse height as a function of drift time for different
values of the drift field for the mixture Xe,CO$_2$(15\%).}
\label{aa:ph3} 
\end{figure} 

In Fig.~\ref{aa:ph3} we show the average pulse height distributions as
a function of drift time for different values of the drift field
for the standard gas mixture, Xe,CO$_2$(15\%).
Beyond the decrease of the pulse height due to longer drift times for lower
drift fields, there is obviously a dramatic loss of signal due to electron 
attachment. 
The relative loss of signal is most pronounced between the two highest values 
of the drift field, for which the variation of the drift velocity is very small.
At drift fields higher than 1~kV/cm, the plateau of the average pulse 
height in the drift region starts to recover.
This trend is explained by the fact that the energy of the drifting 
electrons increases with the electric field and, therefore, the attachment 
coefficient of the SF$_6$-polluted mixture decreases (see next section).
Similar behavior of the attachment as a function of drift field is known
for Ar-based mixtures \cite{aa:huk}.

\begin{figure}[hbt]
\centering\includegraphics[width=.56\textwidth]{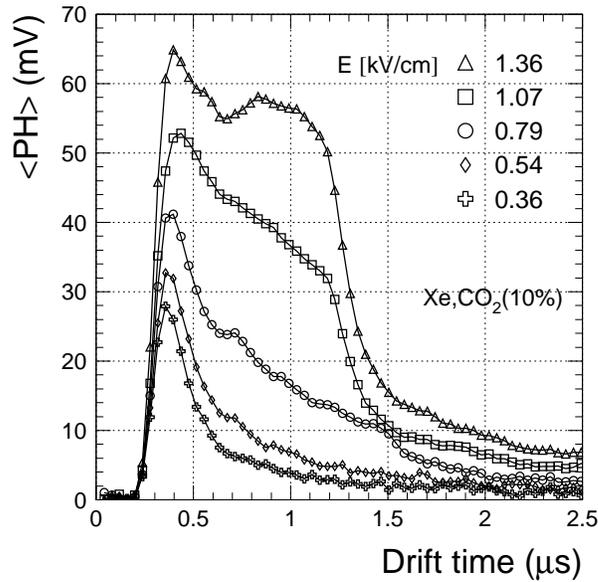}
\caption{As Fig.~\ref{aa:ph3}, but for the mixture Xe,CO$_2$(10\%).}
\label{aa:ph4} 
\end{figure} 

\begin{figure}[hbt]
\centering\includegraphics[width=.56\textwidth]{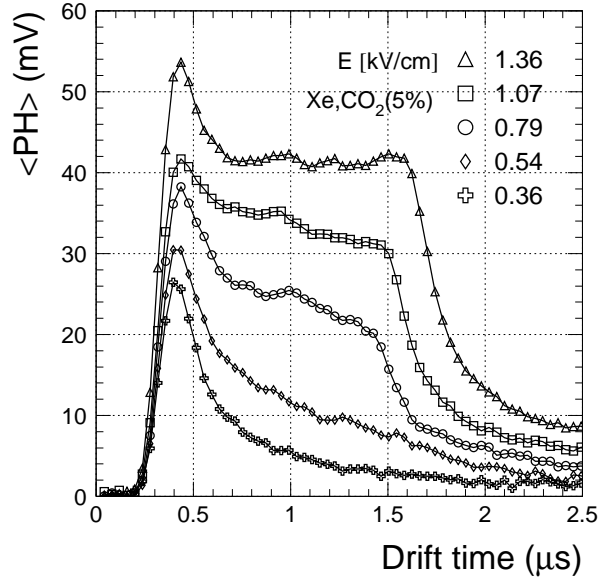}
\caption{As Fig.~\ref{aa:ph3}, but for the mixture Xe,CO$_2$(5\%).}
\label{aa:ph5} 
\end{figure} 

The effect of different CO$_2$ concentrations on the attachment efficiency 
of  the contaminated xenon has also been investigated. 
Fig.~\ref{aa:ph4} and Fig.~\ref{aa:ph5} 
show the average pulse height distributions for 10\% and 5\% CO$_2$, 
respectively.
The gas gains were not identical for the three concentrations of CO$_2$,
so the corresponding distributions can only be compared on a relative 
basis.
Notice that, for the same drift field, the extension of the signal in time
is different for different concentrations of CO$_2$ because of different
drift velocities (see Fig.~\ref{aa:vdrift2} in the next section).
The signal loss due to electron attachment decreases for lower 
CO$_2$ concentrations.
For the Xe,CO$_2$(5\%) mixture, the pulse height distribution almost 
completely recovers at the highest electric field studied here.
As we discuss in the next section, the variation of attachment as a function 
of quencher concentration is due to the dependence of the average energy of
drifting electrons on the CO$_2$ content.

\section{Comparison to calculations} \label{aa:calc}

\begin{figure}[hbt]
\centering\includegraphics[width=.62\textwidth]{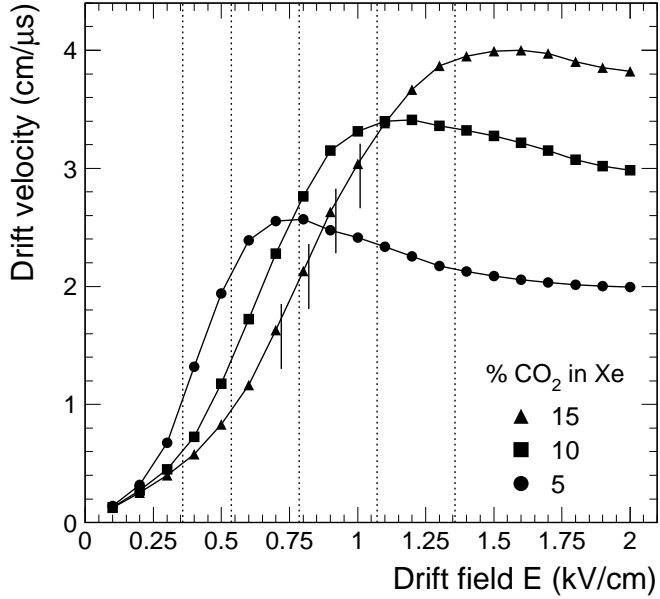}
\caption{Dependence of the drift velocity on the drift field for 5\%, 
10\% and 15\% CO$_2$ content in xenon, as calculated using GARFIELD/MAGBOLTZ 
\cite{aa:garf,aa:magb}.}
\label{aa:vdrift2} 
\end{figure}

To understand the measurements presented above, we have performed calculations 
using the packages GARFIELD  \cite{aa:garf}, MAGBOLTZ \cite{aa:magb} and 
HEED \cite{aa:heed}.
In Fig.~\ref{aa:vdrift2} we present the calculated drift velocities for 
5\%, 10\% and 15\% of CO$_2$ admixture in Xe. 
The dotted vertical lines mark the values of the electric field used for the 
measurements of attachment on SF$_6$. The solid vertical line segments indicate 
the electric fields used for measurements under clean conditions 
(Fig.~\ref{aa:ph1} in section \ref{aa:clean}).
All these values are in a region where the drift velocity has a strong 
dependence on the drift field, a trend reflected in the measurements presented 
above.

\begin{figure}[hbt]
\centering\includegraphics[width=.6\textwidth]{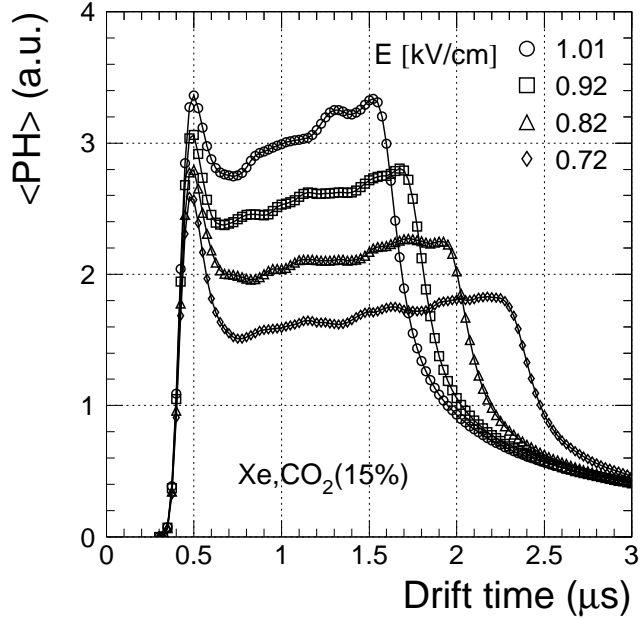}
\caption{Calculated average pulse height as a function of drift time for 
different values of the drift field for the mixture Xe,CO$_2$(15\%),
SF$_6$-free.} \label{aa:phclean} 
\end{figure} 

To compare how well the measured pulse height distributions can be reproduced by 
simulations, we have calculated the detector signals using the GARFIELD package,
under the exact experimental conditions. 
The preamplifier response has been included in these calculations.
The time dependence of the average pulse height for different electric fields 
for the clean Xe,CO$_2$(15\%) mixture is presented in Fig.~\ref{aa:phclean}.
For an easier comparison with the measurements (Fig.~\ref{aa:ph1} in section 
\ref{aa:clean}), we introduced a time shift of 0.4~$\mu$s for these 
distributions. The calculations reproduce the measured signals reasonably well, 
although not in all details.
The slope of signal increase in the drift region as a function of drift time
is larger for the calculations. This may be an indication that some residual
attachment is still present for the measured data, possibly as a result of 
oxygen, water and undetectectable amounts of other contaminants.
The slightly asymmetric gaussian preamplifier response affects this slope 
very little. Already before folding the preamplifier response the calculated 
signals show a larger slope than the measured ones.
For the lower values of the drift field the calculations are in disagreement with
the measurement concerning the time extension of the signal. This discrepancy,
reaching 14\% for the field value of 0.72~kV/cm, may reflect a different field
dependence of the drift velocity in measurements and calculations.
We note that a good agreement was found between calculations and measurements
in other Xe-based mixtures \cite{aa:kunst,aa:becker}.
Since the bow of the entrance window (which is also the drift electrode) due 
to gas overpressure introduces an obvious uncertainty in the present 
measurements, it is too early to assess the above discrepacy quantitatively. 
Precision measurements of the drift velocity for the standard TRD gas mixture 
Xe,CO$_2$(15\%) are in progress.

\begin{figure}[hbt]
\centering\includegraphics[width=.6\textwidth]{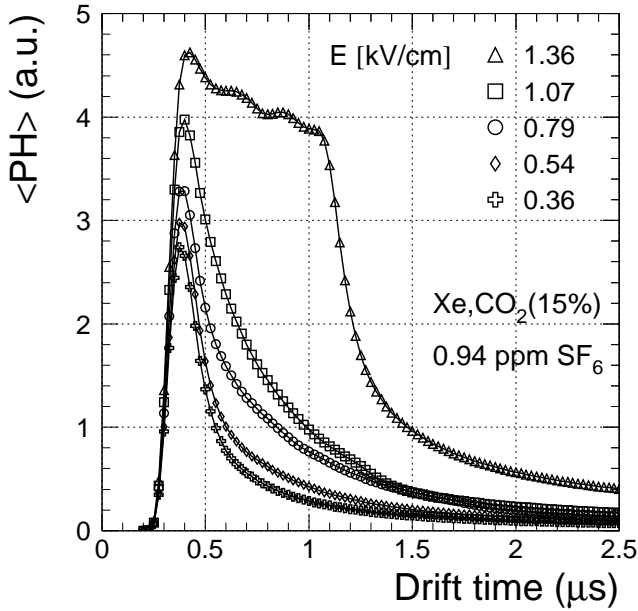}
\caption{Calculated average pulse height as a function of drift time for 
different values of the drift field for the mixture Xe,CO$_2$(15\%),
with 0.94~ppm of SF$_6$ contamination.}
\label{aa:phsf6} 
\end{figure} 

Figure~\ref{aa:phsf6} shows the time dependence of the average pulse height
for different electric fields for the Xe,CO$_2$(15\%) mixture with 0.94~ppm 
of SF$_6$, corresponding to the SF$_6$ fraction in the measurements.
For a direct comparison with the measurements, the time shift is 0.3~$\mu$s 
in this case.
The measured attachment (Fig.~\ref{aa:ph3} in the previous section) is 
reproduced almost quantitatively, except for the smoother reduction of 
attachment towards higher field values seen in the measurements.
Not reproduced is the measured stronger variation of the signal in the
amplification region.

\begin{figure}[hbt]
\centering\includegraphics[width=.6\textwidth]{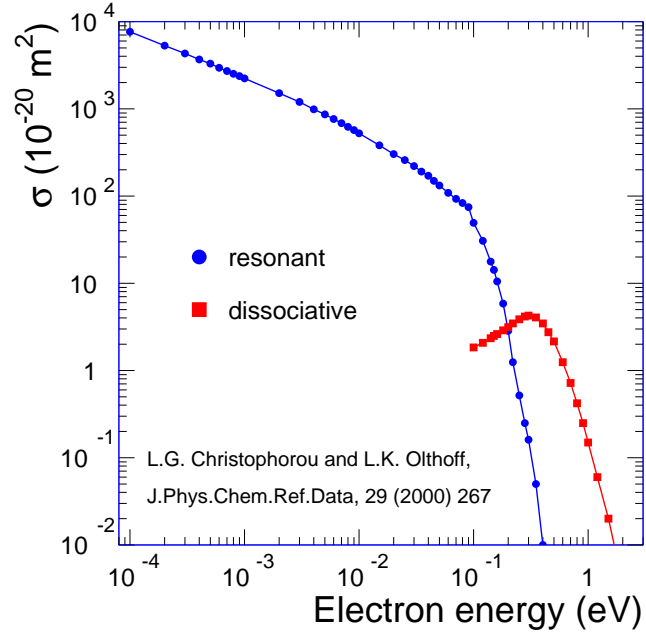}
\caption{The energy dependence of cross section of electron attachment on SF$_6$.
Recommended values from ref. \cite{aa:chris} are plotted.}
\label{aa:att-sigma} 
\end{figure}

We turn now to a more detailed investigation of the attachment on SF$_6$.
In particular, the measured drift field dependence of the attachment (nicely 
reproduced by simulations) is expected to reflect the characteristics 
of the attachment cross section as a function of electron energy.
Indeed, the cross section of electron attachment on SF$_6$ is large and has a 
strong energy dependence, as shown in in Fig.~\ref{aa:att-sigma} \cite{aa:chris}.
Resonance capture, which leads to the formation of SF$_6^-$, is the most 
important mechanism, with cross sections up to 8$\cdot$10$^{-17}$~m$^2$ for 
near-zero electron energies.
The cross section of dissociative capture, resulting in a free fluorine atom
and SF$_5^-$, peaks at the value of 4$\cdot$10$^{-20}$~m$^2$ for electron 
energy of about 0.3~eV.  
These characteristics of attachment cross sections of SF$_6$ are quite 
different compared to the O$_2$ case \cite{aa:chris2}.
For example, in case of O$_2$, the cross section for dissociative capture peaks
at about 10$^{-22}$~m$^2$ for electron energy of 6.5~eV \cite{aa:chris2}.

\begin{figure}[hbt]
\begin{tabular}{lr}\begin{minipage}{.48\textwidth}
\centering\includegraphics[width=1.0\textwidth,height=1.08\textwidth]{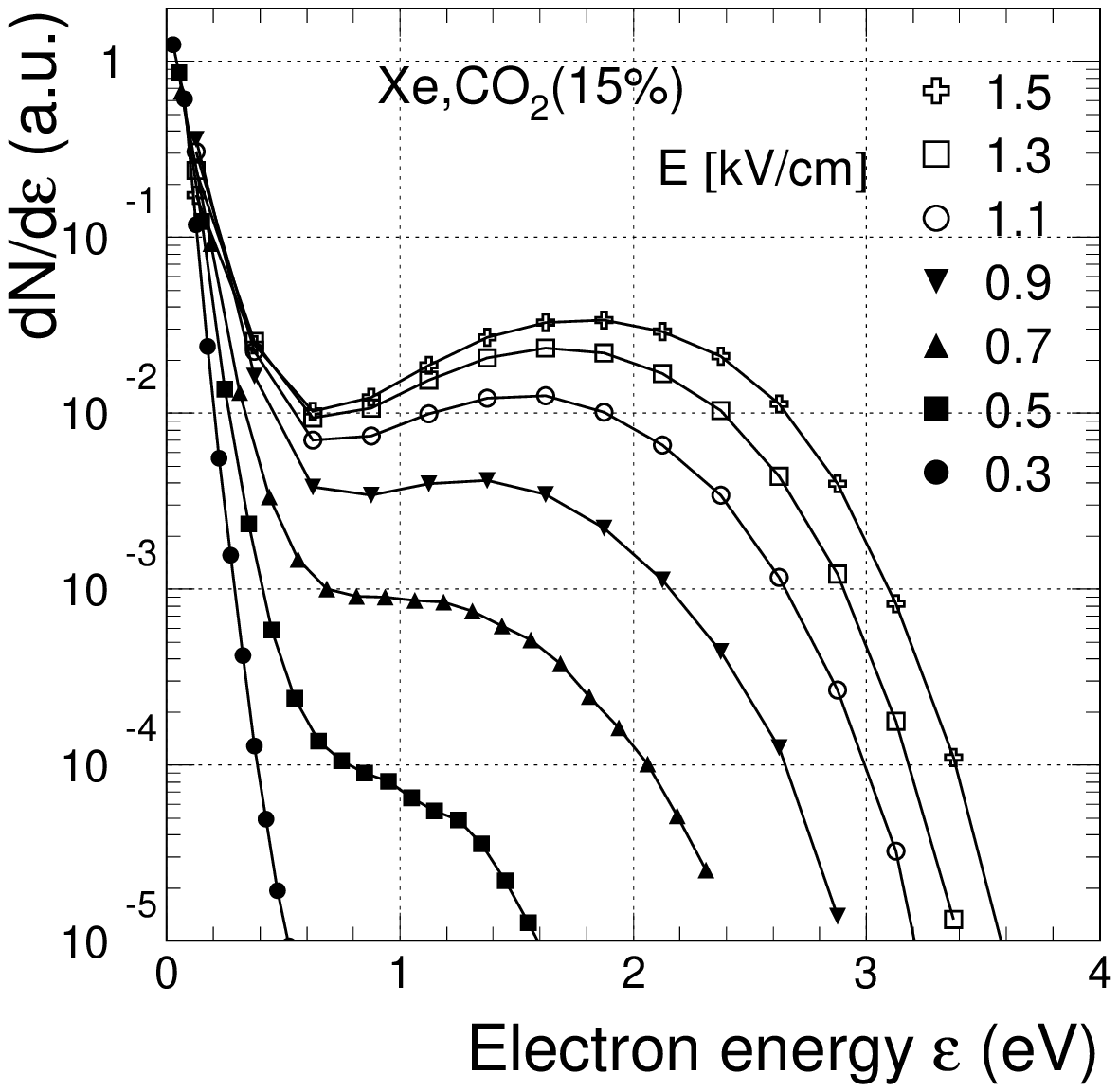}
\caption{Energy distribution of drifting electrons for different values of 
the drift field for the mixture Xe,CO$_2$(15\%).}
\label{aa:endis} 
\end{minipage} & \begin{minipage}{.48\textwidth}
\centering\includegraphics[width=1.0\textwidth,height=1.08\textwidth]{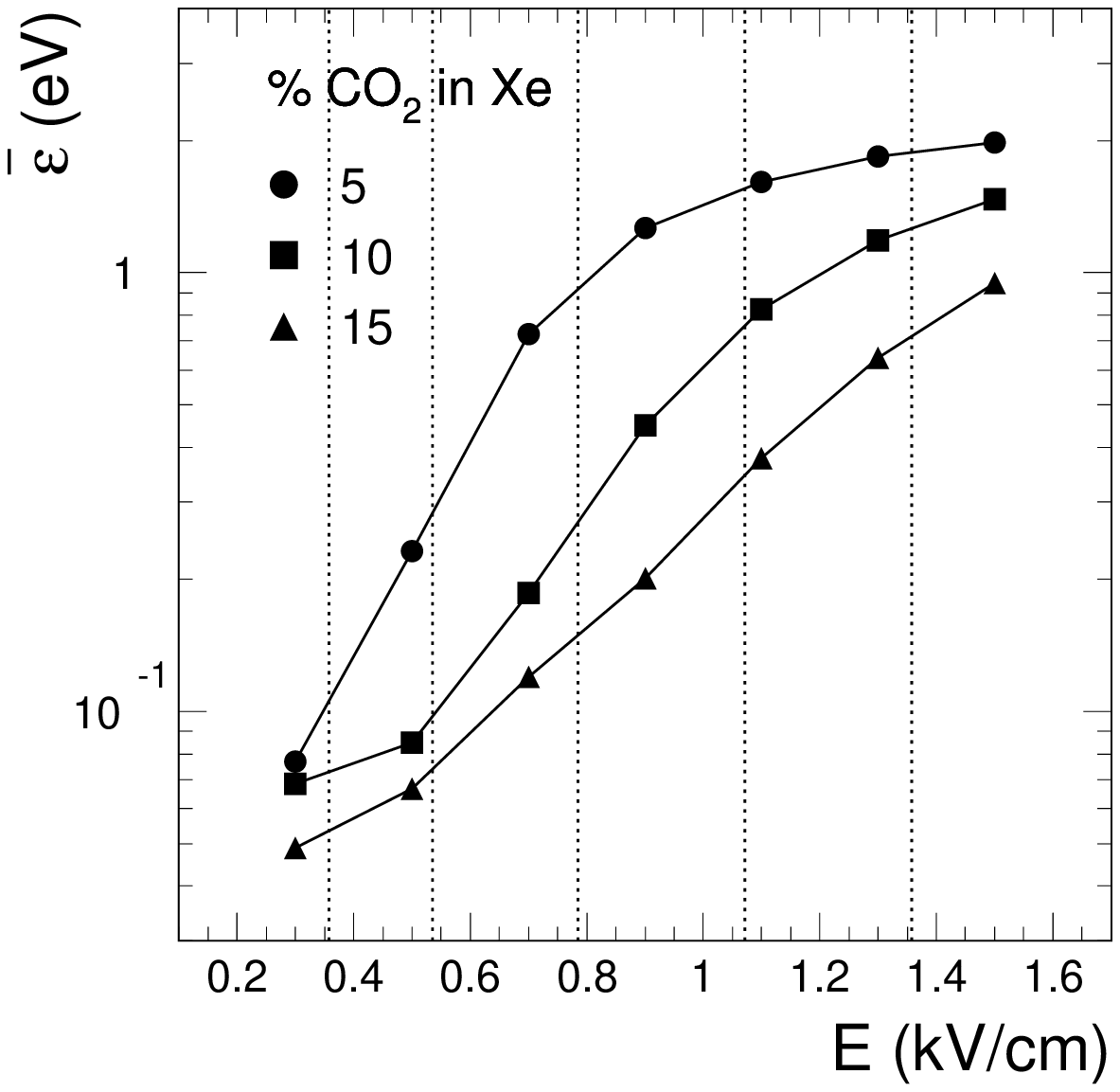}
\caption{Average energy of drifting electrons as a function of the drift field 
for different concentrations of CO$_2$ in Xe.}
\label{aa:emean} 
\end{minipage} \end{tabular}
\end{figure}

The energy spectra of the drifting electrons for various electric fields are 
shown in Fig.~\ref{aa:endis} for the mixture Xe,CO$_2$(15\%).
These distributions are computed with the simulation program Imonte 
\cite{aa:magb}.
A significant high-energy component (energies above 1~eV) is present towards
higher field values.
The average energy of drifting electrons (also computed with Imonte), 
$\bar{\varepsilon}$, is plotted in Fig.~\ref{aa:emean} as a function of drift 
field for the three concentrations of CO$_2$ in Xe. 
The average energy increases strongly with the electric field and is larger
for smaller CO$_2$ concentrations.
The dotted vertical lines mark the values of the drift field used for the 
measurements.
As the attachment cross section decreases as a function of the electron energy, 
the trends seen in Fig.~\ref{aa:emean} explain both the observed dependence of 
attachment on drift field and on CO$_2$ concentration mixture.

The concentration of CO$_2$ may influence the rate of the resonant energy
transfer in the second step of the Bloch-Bradbury process (\ref{aa:eq2}),
but this seems  not to be the case for attachment on SF$_6$.
The measured pulse height distributions for 15\%, 10\% and 5\% CO$_2$ 
(Figs.~\ref{aa:ph3} to \ref{aa:ph5} in the previous section) for the field 
values of 1.36, 1.07 and 0.79 kV/cm (for which the corresponding average 
energies are similar, see Fig.~\ref{aa:emean}), respectively, look very similar.
Although we cannot make a quantitative statement, apparently all the influence 
of CO$_2$ concentration on the attachment stems from the average energy of 
the drifting electrons.
In contrast, a dependence of oxygen attachment on quencher content 
(beyond electron energy contribution) has been measured for 10\% and 20\% CH$_4$ 
in Ar \cite{aa:huk}.
This difference between SF$_6$ and O$_2$ attachment is probably the result of 
different lifetimes of the corresponding excited states of the negative ions.
Indeed, the autodetachment lifetime is larger than 1~$\mu$s for SF$_6^{*-}$ 
\cite{aa:chris2}, comparable with the drift time in our detectors, 
whereas for O$_2^{*-}$ it is about 10$^{-4}$~$\mu$s \cite{aa:kok}.

\section{Methods for checking the gas quality} \label{aa:check}

Using xenon supplies as SF$_6$-free as possible is an important requirement. 
In the following we describe our monitoring procedures, i.e. gas chromatography
and $^{55}$Fe pulse height measurements using dedicated monitor detectors.

Gas chromatography allows the detection and quantification of traces of 
pollutants in a gas, by separating the different species contained in the 
sampled gas in a chromatograph column, and by detecting them in a suitable 
detector placed downstream of the column. The separated effluents give rise to 
characteristic peaks in a time diagram which can be identified and quantified 
after proper calibration of the device.
Since SF$_6$ has a high electron capture cross section, an Electron 
Capture Device (ECD) is a suitable detector. An ECD consists of a cavity 
through which a so-called make-up gas (nitrogen) flows. 
Electrons from a beta source ($^{63}$Ni, maximum energy 66 keV) partly 
ionise the nitrogen gas. The total current produced is collected 
by an electrode. 
If an electronegative substance flows through the cavity at a given time, 
the missing collected charge is converted into a
peak in the corresponding chromatogram. The area under the peak is proportional 
to the amount of electrons captured.

\begin{figure}[hbt]
\centering\includegraphics[width=.68\textwidth]{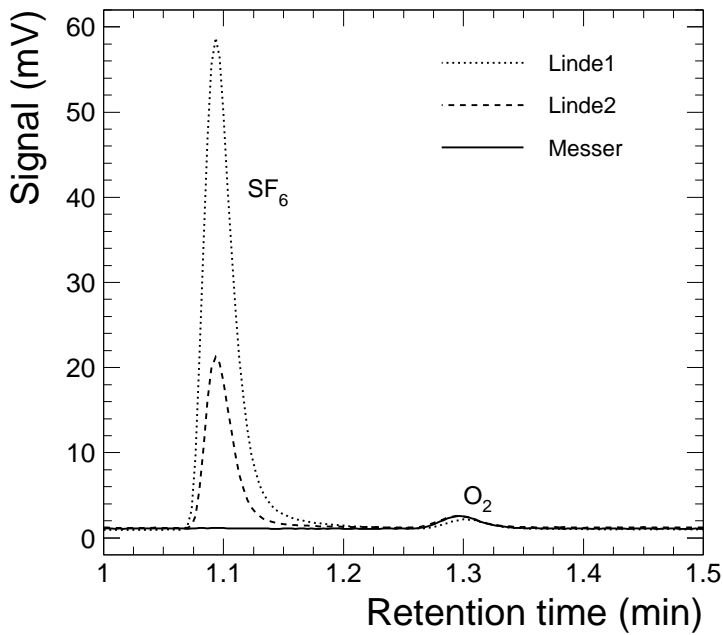}
\caption{Chromatograms of syringe-injected samples of three xenon supplies: 
SF$_6$-free (Messer) and with SF$_6$ contaminations of 0.3 ppm (Linde2) and 
1.1 ppm (Linde1).}
\label{aa:chro} 
\end{figure} 

An ECD has been connected to our gas chromatograph \cite{aa:finn} 
in order to analyse our different supplies of xenon. 
Concentrations as low as 1~ppb are detectable.
The device has been calibrated for SF$_6$ by putting small amounts of this 
gas into a glass container filled with helium. Special care has to be 
taken with the injection syringes since they get temporarily contaminated 
when  exposed to high concentrations of SF$_6$.
In addition to the two xenon supplies (Messer and Linde1) used for the beam 
measurements reported above (section \ref{aa:sf6}), we have also investigated 
a more recent supply from Linde (Linde2).
The resulting chromatograms for the two kinds of xenon from Linde and the xenon 
from Messer-Griesheim are shown in Fig.~\ref{aa:chro}. 
The calibration yields 1.1 and 0.35 ppm SF$_6$ for the Linde1 and Linde2 gas 
samples, respectively, with an error of about 15\%. The xenon from 
Messer-Griesheim showed 1.5 ppb SF$_6$, most probably coming from the 
contaminated syringe.
The injection method also leads to inevitable air contamination, as revealed by 
the oxygen peak (corresponding to 400 ppm) in the chromatograms.

\begin{figure}[hbt]
\centering\includegraphics[width=.57\textwidth]{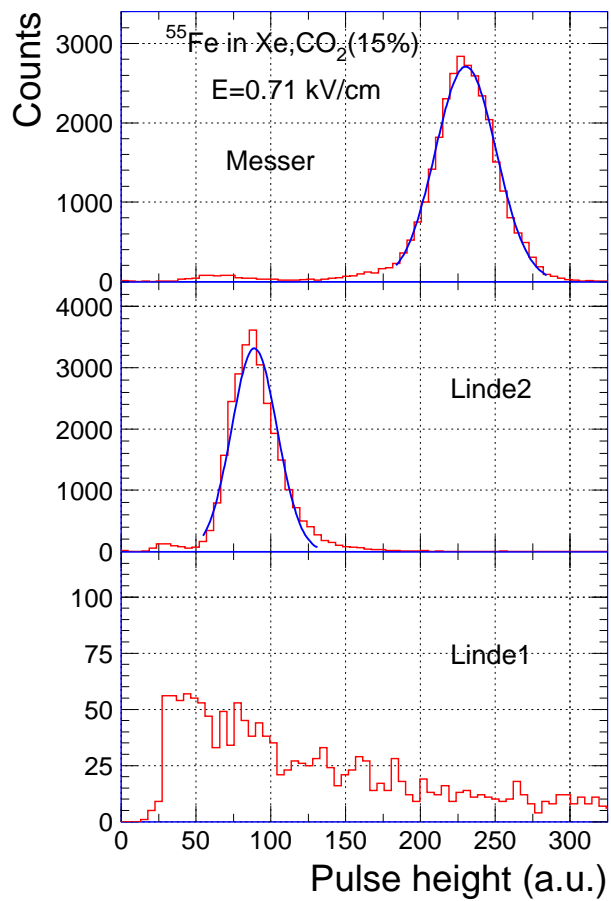}
\caption{Comparison of pulse height distributions of $^{55}$Fe in 
Xe,CO$_2$(15\%) for three supplies of xenon.}
\label{aa:fe} 
\end{figure} 

Gas chromatography is a simple, accurate and economic 
technique, but the chromatograph itself is a rather
expensive device. However, for most detector applications, one does not
necessarily need chromatography
in order to assess the gas quality. Measurements with a $^{55}$Fe source on 
standard detectors are sensitive enough to reveal possible problems due to
attachment.
In our case, in order to minimize Xe consumption for such tests, we have 
built small monitor detectors for the special purpose of checking 
the gas quality using $^{55}$Fe pulse height measurements.
These monitor detectors have the same electrical field configuration as the 
drift chambers used for the beam measurements described above.
Their small volume of about half a liter minimizes the loss of xenon gas. 
A collimated $^{55}$Fe source is placed in front of the entrance window. 
As the X-rays are absorbed preferentially at the beginning of the drift 
region, the cluster of primary electrons drifts in most cases 3~cm and is 
subject to attachment over this distance.
For these measurements the anode voltage has been tuned for a gas gain of about 
10$^4$. 
Here we have used a different preamplifier, with a gain of 6~mV/fC and 
noise of about 1000 electrons r.m.s.
As a result of a low flow through the detector, the contamination with O$_2$ 
and H$_2$O was 70 and 400~ppm, respectively.

Figure~\ref{aa:fe} shows pulse height spectra for the three supplies of xenon 
discussed above.
The value of the drift field (which corresponds roughly to the 
anticipated operational point of the final TRD in ALICE) is 0.71 kV/cm.
Compared to clean Xe (Messer), the 0.35 ppm SF$_6$ contamination in Linde2 
Xe leads to a pulse height distribution with a much smaller value for the 
main peak. 
In addition, a clear tail towards larger pulse height is seen, originating from
absorption of X-rays deeper into the drift region, and thereby subject to
less attachment loss.
In case of Linde1 (1.1~ppm SF$_6$) the $^{55}$Fe signal is completely
lost. The spectrum recorded is the result of X-rays absorbed in the vicinity 
of the anode and of cosmic-ray background. 
Notice that the number of counts is much smaller in this case
(for a comparable acquisition time).
Notice also that, for the Messer and Linde2 cases, the escape peak of Xe 
(at 1.76 keV, compared to 5.96~keV full energy of $^{55}$Fe)
is clearly visible 
(in case of Linde2, the escape peak is partially cut by the threshold).

\begin{figure}[hbt]
\centering\includegraphics[width=.55\textwidth]{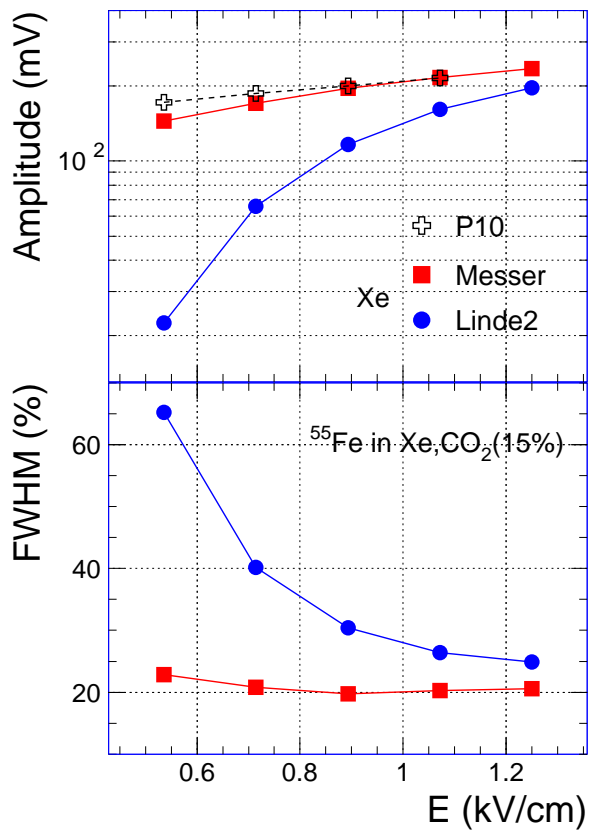}
\caption{Centroid (upper panel) and energy resolution (lower panel) 
of pulse height distributions of $^{55}$Fe spectra in Xe,CO$_2$(15\%) 
as a function of drift field for two supplies of xenon.}
\label{aa:ed} 
\end{figure} 

When possible, we performed gaussian fits of the main peak (also plotted in 
Fig.~\ref{aa:fe}) and extracted the centroid and energy resolution. 
Both quantities are influenced by attachment.
In Fig.~\ref{aa:ed} we show the dependence of the amplitude of the main peak
and its FWHM on the drift field for Messer and Linde2 supplies. 
The clean gas shows the expected variation of amplitude as a function 
of drift field arising from gain increase due to the transparency of the cathode 
wire grid. 
For comparison we include a measurement with an Ar,CH$_4$(10\%) mixture 
(P10). Compared to this, at low fields, even for clean Xe, small 
deviations are seen, consistent with attachment on O$_2$; taking the
P10 signal as reference, 
we deduce an attachment coefficient of 506 bar$^{-2}\mu$s$^{-1}$ for the 
field value of 0.72 kV/cm, in reasonable agreement to the value determined
from beam tests (section~\ref{aa:oxygen}).
For the Linde2 case this dependence of the signal on electric field is much 
stronger, as a result of the attachment.
This is confirmed by the energy resolution (lower panel in Fig.~\ref{aa:ed}), 
which improves dramatically when going (as a function of drift field) from 
strong to weak attachment.

\begin{figure}[hbt]
\centering\includegraphics[width=.62\textwidth]{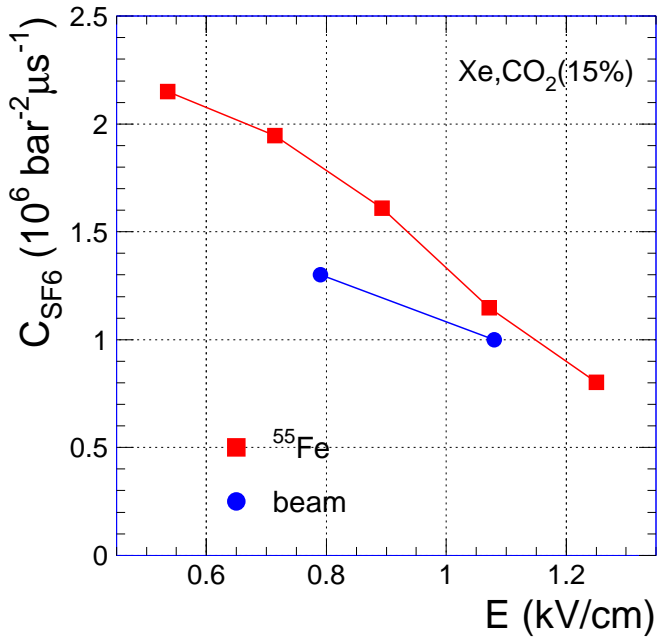}
\caption{Attachment coefficient on SF$_6$ in Xe,CO$_2$(15\%) as a function of 
drift field.}
\label{aa:atcoef} 
\end{figure} 

We have determined the attachment coefficient on SF$_6$ (under our experimental 
conditions mentioned above) using the ratio of the $^{55}$Fe amplitudes for the 
Messer and Linde2 cases. The drift time is extracted using the drift velocities 
calculated with GARFIELD and assuming that the absorption of X-rays takes place 
right at the entrance window.
The results are presented in Fig.~\ref{aa:atcoef} as a function of drift field.
As expected, the attachment coefficient on SF$_6$ is large, of the order of 
10$^6$ bar$^{-2}\mu$s$^{-1}$.
In addition, it has a pronounced dependence on the drift field, in line with our
arguments presented in section \ref{aa:calc}.
If we consider the absorption length of 5.9~keV X-rays in the Xe,CO$_2$(15\%) 
mixture, which is about 3 mm, the attachment coefficient would increase by 10\%.
For comparison, we include in this plot the results obtained from the beam 
mesurements, taking as a reference the time distribution of signals measured 
in the clean case and normalized in the amplification region. 
Taking into account the uncertainties, in particular coming from the 
normalization for the beam measurements, the extracted values for the attachment
coefficient are in reasonable agreement.

\section{Summary} \label{aa:conc}

We have performed measurements of pulse height distributions in drift chambers
operated with Xe,CO$_2$ mixtures.
After studying the general behavior of these distribution under clean 
conditions, we have investigated the role of oxygen and SF$_6$ contamination 
of the detection gas. 
A small signal loss due to attachment is seen for O$_2$ impurities 
up to a few hundred ppm. 
In case of SF$_6$, a contamination even at the level below 1~ppm 
produces a dramatic loss of signal over our drift length of about 3~cm.
Attachment on SF$_6$ is studied here for the first time 
concerning its practical implications for gas detectors.
As the SF$_6$ was found accidentally in some xenon supplies, it is important to
have a careful monitoring of the SF$_6$ contamination when precision measurements
are performed using Xe-based gas mixtures in drift chambers.
We have used ECD gas chromatography analysis to detect and quantify small traces 
of SF$_6$.
We have shown that measurements of $^{55}$Fe signals in monitor detectors are 
very sensitive to SF$_6$ contamination, thus allowing an inexpensive in situ 
check of the gas quality.

\section*{Acknowledgments}
We acknowledge the skills and dedication of A. Radu and J. Hehner in building
our detectors.
We are indebted to S. Ilie and C. Jeanpetit for the first analysis of our gas.
We appreciate the advice from R. Veenhof concerning the GARFIELD calculations.
We acknowledge P. Szymanski for help during the measurements at CERN.

\end{document}